\global\def\draftcontrol{0}

   \def\versionno{ alpha prime t11 -- draft   }

\catcode`\@=11

\expandafter\ifx\csname draftcontrol\endcsname\relax\global\def\draftcontrol{0}
\fi

{\count255=\time\divide\count255 by 60
\xdef\hourmin{\number\count255}
\multiply\count255 by-60\advance\count255 by\time
\xdef\hourmin{\hourmin:\ifnum\count255<10 0\fi\the\count255}}
\def\draftdate{\number\month/\number\day/\number\year\ \ \ \hourmin }

\newcommand\makepapertitle{\par
  \begingroup
    \renewcommand\thefootnote{\@fnsymbol\c@footnote}%
    \def\@makefnmark{\rlap{\@textsuperscript{\normalfont\@thefnmark}}}%
    \long\def\@makefntext##1{\parindent 1em\noindent
            \hb@xt@1.8em{%
                \hss\@textsuperscript{\normalfont\@thefnmark}}##1}%
     \newpage
     \global\@topnum\z@   
     \@makepapertitle
     \thispagestyle{empty}\@thanks
  \endgroup
  \setcounter{footnote}{0}%
  \global\let\thanks\relax
  \global\let\makepapertitle\relax
  \global\let\@makepapertitle\relax
  \global\let\@thanks\@empty
  \global\let\@author\@empty
  \global\let\@date\@empty
  \global\let\@title\@empty
  \global\let\title\relax
  \global\let\author\relax
  \global\let\date\relax
  \global\let\and\relax
  \def\version{\let\version\@version\@gobble}
}
\def\@makepapertitle{%
  \newpage
   \ifnum\draftcontrol=1 {}
   \version\versionno
   \vskip 3em%
   \else
   \hfill\hbox to 3cm {\parbox{4cm}{\@pubnum}\hss}%
   \vskip 3em%
   \fi
   \begin{center}%
   \let \footnote \thanks
     {\LARGE {\@title}}%
     \vskip 1.5em%
     {\normalsize
       \lineskip .5em%
       \begin{tabular}[t]{c}%
         \@author
       \end{tabular}\par}%
     \vskip 1.5em%
     {\@bstract}%
     \end{center}%
     \vskip 1.5em
     \@date%
   \par
}

\gdef\@pubnum{}
\def\pubnum#1{%
  \gdef\@pubnum{#1}}

\gdef\@bstract{}
\def\Abstract#1{%
  \gdef\@bstract{%
   \parbox{\textwidth-0pc}{%
   \centerline{\bf Abstract}\penalty1000%
\kern.2cm%
\noindent
\renewcommand\baselinestretch{1.0}%
{#1}}}
}

\def\ps@paper{\let\@mkboth\@gobbletwo%
     \ifnum\draftcontrol=1
	\def\@oddfoot{\hbox to \textwidth{\tiny \versionno \hfil\tiny\draftdate}%
	\hskip -\textwidth \hbox to \textwidth{\hfil\rm\thepage\hfil}}%
     \else\def\@oddfoot{\hbox to \textwidth{\hfil\rm\thepage\hfil}}
     \fi
     \let\@evenfoot\@oddfoot
}

\def\body{\clearpage
          \pagestyle{paper}
	}

\def\@version#1{\ifnum\draftcontrol=1
\typeout{}\typeout{#1}\typeout{}
\vskip3mm\centerline{\hbox{\fbox{\normalsize{\tt DRAFT -- #1 -- }
                   {\draftdate}}}}\vskip3mm
\fi}
\let\version\@version
\long\def\eqlabel#1{\ifnum\draftcontrol=1
                    \tag@false  
                    \tag*{(\theequation) \hbox to -0.2cm{\hspace{0cm}\small{#1}\hss}}
                    \refstepcounter{equation}
                    \edef\@currentlabel{\theequation}
                    \ltx@label{#1}          
                    \else
                    \label{#1}
                    \fi
                    }
\let\st@bibitem\@bibitem
\let\st@lbibitem\@lbibitem
\ifnum\draftcontrol=1
  \def\@bibitem#1{%
    \st@bibitem{#1}\a@@label{#1}\ignorespaces}
  \def\@lbibitem[#1]#2{%
    \st@lbibitem[#1]{#2}\a@@label{#2}\ignorespaces}
  \def\a@@label#1{%
    \gdef\a@lab{\smash{\normalfont\small#1}}
    \ifvmode
      \if@inlabel
        \global\setbox\@labels\hbox{%
          \llap{\a@lab\let\a@lab\relax
                \kern\@totalleftmargin\kern\marginparsep}%
          \box\@labels}%
      \fi
    \fi}
\fi

\documentclass[12pt,letterpaper]{article}

\usepackage{amsmath,amssymb,array,calc,rotating,epsfig,psfrag}
\usepackage[nosort]{cite}

\ifnum\draftcontrol=1
\fi

\renewcommand\baselinestretch{1.25}
\renewcommand\section{\@startsection {section}{1}{\z@}%
                                   {-3.5ex \@plus -1ex \@minus -.2ex}%
                                   {2.3ex \@plus.2ex}%
                                   {\normalfont\large\bfseries}}
\renewcommand\subsection{\@startsection{subsection}{2}{\z@}%
                                   {-3.25ex\@plus -1ex \@minus -.2ex}%
                                   {1.5ex \@plus .2ex}%
                                   {\normalfont\normalsize\bfseries}}
\renewcommand\subsubsection{\@startsection{subsubsection}{3}{\z@}%
                                   {-3.25ex\@plus -1ex \@minus -.2ex}%
                                   {1.5ex \@plus .2ex}%
                                   {\normalfont\normalsize\it}}
\renewcommand\paragraph{\@startsection{paragraph}{4}{\z@}%
                                   {-3.25ex\@plus -1ex \@minus -.2ex}%
                                   {1.5ex \@plus .2ex}%
                                   {\normalfont\normalsize\bf}}

\numberwithin{equation}{section}

\def\ie{{\it i.e.}}

\def\revise#1       {\raisebox{-0em}{\rule{3pt}{1em}}%
                     \marginpar{\raisebox{.5em}{\vrule width3pt\
                     \vrule width0pt height 0pt depth0.5em
                     \hbox to 0cm{\hspace{0cm}{%
                     \parbox[t]{4em}{\raggedright\footnotesize{#1}}}\hss}}}}

\def\calc         {{\cal C}}

\def\calf         {{\cal F}}

\def\caln         {{\cal N}}
\def\calo         {{\cal O}}

\def\sqr#1#2{{\vcenter{\vbox{\hrule height.#2pt
 \hbox{\vrule width.#2pt height#1pt \kern#1pt
 \vrule width.#2pt}\hrule height.#2pt}}}}

\newcommand{\ft}[2]{{\textstyle{\frac{#1}{#2}}}}

\def\a{\alpha}
\def\w{\omega}
\def\g{\gamma}

\newcommand{\qq}{\mathfrak{q}}
\newcommand{\ww}{\mathfrak{w}}

\catcode`\@=12

\begin{document}


\title{Shear viscosity of CFT plasma  
at finite coupling}

\pubnum{%
UWO-TH-08/7
}
\date{April 2008}

\author{
Alex Buchel\\[0.4cm]
\it Department of Applied Mathematics\\
\it University of Western Ontario\\
\it London, Ontario N6A 5B7, Canada\\[0.2cm]
\it Perimeter Institute for Theoretical Physics\\
\it Waterloo, Ontario N2J 2W9, Canada\\
}

\Abstract{
We present evidence for the universality of the shear viscosity of
conformal gauge theory plasmas beyond infinite coupling. We comment of
subtleties of computing the shear viscosity in effective models of
gauge/gravity correspondence rather than in string theory.
}


\makepapertitle

\body

\version\versionno

\section{Introduction}
Various universal features of transport properties of strong coupled gauge theory plasmas 
where discovered within the framework of gauge theory/string theory correspondence of Maldacena \cite{m9711,m2}.
For example, all gauge theory plasmas (various gauge groups, matter content, with or without 
chemical potentials for conserved $U(1)$ charges, with non-commutative spatial directions) that allow for a dual supergravity description
have a universal value of the shear viscosity at infinite 't Hooft coupling \cite{u1,u2,u3,u4,u5,u55} . 
Similarly, while not universal, the bulk viscosity of non-conformal  gauge theory plasmas 
(again at infinitely strong coupling and provided the dual holographic description is available)
appear to satisfy a universal bound \cite{u6}. 

Much less is known about viscosity of gauge theory plasma at finite coupling:
the shear viscosity of $\caln=4$ supersymmetric Yang-Mills plasma at finite coupling 
was computed in \cite{f1,f2,f3,f4}\footnote{It was also computed  in some phenomenological models of 
gauge/gravity correspondence in \cite{p1,p2}.}. The difficulties of extending the analysis of 
\cite{f1,f2,f3,f4} to other examples of gauge/string correspondence are both technical and conceptual. 
On a technical side, one deals with a daunting task of studying quasinormal modes of near-extremal 
black branes in full ten dimensional type IIB supergravity, including (at least to leading order) 
higher derivative $\calo(\a'^3)$-corrections coming from integrating out massive modes of type IIB string.  
Conceptually, the problem is that the full set of such $\calo(\a'^3)$-corrections is not known presently.
Typically, in studies of black brane thermodynamics \cite{gkt1,pt1,th1} and hydrodynamics \cite{f1,f2,f3,f4}
at $\calo(\a'^3)$ one takes into account only curvature corrections to type IIB supergravity 
\cite{cor1,cor2,cor3,cor4}. 
What is known is that additional terms must be present \cite{hss,pt,p} and that they 
might affect the hydrodynamics of dual gauge theory plasma.
Thus, the available results \cite{f1,f3} for the shear viscosity of $\caln=4$ SYM plasma can 
receive further corrections from Ramond-Ramond fluxes in the dual string theory description. 
We emphasize though that such corrections will be additive (to leading order in $\a'$), 
and contribution of  $\calo(\a'^3)$ terms  due to fluxes can be computed on top of type IIB supergravity, 
excluding purely higher-derivative curvature corrections.  Here, we consider only 
higher-derivative curvature corrections to type IIB supergravity --- as a result our conclusions 
are subject to caveat alluded to above.

In this letter we present evidence for the universal features of conformal gauge theory plasmas at finite 
't Hooft coupling. A claim of this type can make sense only if a prescription is given as to how to compare 
different gauge theories\footnote{This was not the issue in either shear viscosity universality 
or bulk viscosity bound at infinite coupling.}. For example, for $SU(N_c)$ $\
\caln=4$ SYM plasma at 't Hooft coupling $\lambda_{\caln=4}$ 
\begin{equation}
\frac{\a'}{L^2}=\frac{1}{\sqrt{\lambda_{\caln=4}}}\,,
\eqlabel{ln4} 
\end{equation}
where $L$ is the radius of $AdS_5$.
Now, for a general CFT (with a dual string theory description) we have \cite{lrw}
\begin{equation}
\frac{\a'}{L^2}=\frac{1}{\sqrt{\lambda_{CFT}}}\ \sqrt{\frac{a_{\caln=4}}{a_{CFT}}}\,,
\eqlabel{lcft} 
\end{equation}
where $a_{\caln=4}$ and $a_{CFT}$ are the central charges of the $\caln=4$ SYM and a given CFT
correspondingly. We identify  two CFT gauge theory plasmas if according to \eqref{lcft} they have the 
same holographic string tension, \ie, 
\begin{equation}
\lambda_{CFT_1}\ a_{CFT_1}=\lambda_{CFT_2}\ a_{CFT_2}\,.
\eqlabel{iden}
\end{equation}
Our evidence for the universality of the shear viscosity of CFT plasma at finite coupling comes from 
analysis of a shear viscosity of $\caln=1$ superconformal gauge theory plasma of Klebanov-Witten (KW)
\cite{kw}. Specifically, we show that the full spectrum of shear quasinormal modes of KW plasma
at order $\calo(\a'^3)\sim \calo\left(\lambda_{KW}^{-3/2}\right)$  is identical to that of the $\caln=4$ plasma at order
$\calo\left(\lambda_{\caln=4}^{-3/2}\right)$. 
For a specific case of  the lowest quasinormal modes, this immediately implies 
that the ratio of the shear viscosity to the entropy density is the same in KW and $\caln=4$ SYM plasmas.
Note that following our convention \eqref{iden}, we set
\begin{equation}
\frac{\lambda_{KW}}{\lambda_{\caln=4}}=\frac{16}{27}\,.
\end{equation} 

Gauge theory plasma transport analysis in \cite{f1,f2,f3} and in this letter are done in the 
the full ten dimensional type IIB supergravity including higher-derivative curvature corrections 
\cite{cor1,cor2,cor3,cor4}. As a separate computation, we show that similar analysis done in lower
dimensional higher-derivative effective action (obtained without proper Kaluza-Klein reduction)
could be misleading. While a specific example we present gives the same value of the shear viscosity 
as in its higher dimensional counterpart, the agreement is accidental and the two models
have different spectra of quasinormal modes as well as different higher order hydrodynamic corrections 
for the dispersion relation of the lowest quasinormal mode.   
In view of this, it would be interesting to re-examine the claims in \cite{p1,p2}.

\section{KW plasma at finite temperature and 't Hooft coupling}
A string theory dual to KW gauge theory is a near-horizon limit of D3 brane on the conifold \cite{kw}.
We begin with computing leading order $\a'$ corrections to near-extremal D3 branes at the conifold, dual 
to finite temperature KW plasma at order $\calo\left(\lambda^{-3/2}\right)$ in the 't Hooft coupling.

Type IIB supergravity effective action with only higher-derivative curvature corrections takes form
\begin{equation}
I=  \frac{1}{ 16\pi G_{10}} \int d^{10} x \sqrt {-g}
\ \bigg[ R - {\frac 12} (\partial \phi)^2 - \frac{1}{4 \cdot 5!}   (F_5)^2  +...+ 
\  \gamma \ e^{- {\frac 3 2} \phi}  W + ...\bigg]   \  ,
\eqlabel{aaa}
\end{equation}
$$ \ \ \ \ \ \   
  \gamma= { \frac 18} \zeta(3)(\alpha')^3 \ , 
$$  
where 
\begin{equation}
W =  C^{hmnk} C_{pmnq} C_{h}^{\ rsp} C^{q}_{\ rsk} 
 + {\frac 12}  C^{hkmn} C_{pqmn} C_h^{\ rsp} C^{q}_{\ rsk}\  . 
\eqlabel{rrrr}
\end{equation}
 In \eqref{aaa} ellipses stand for $\calo(\a'^3)$ contribution of Ramond-Ramond fluxes that we neglect here.

We represent  ten dimensional background geometry describing $\gamma$-corrected black 3-branes on the conifold 
by the following ansatz
\begin{equation}
\begin{split}
ds_{10}^2=&g_{\mu\nu}^{(0)}\ dx^\mu dx^\nu+c_4^2 e_\psi^2+c_5^2\sum_{a=1}^2
\left(e_{\theta_a}^2+e_{\phi_a}^2\right)\\
\equiv&-c_1^2 dt^2+c_2^2\left(dx^2+dy^2+dz^2\right)+c_3^2dr^2  +c_4^2 e_\psi^2+c_5^2\sum_{a=1}^2
\left(e_{\theta_a}^2+e_{\phi_a}^2\right)\,,
\end{split}
\eqlabel{g10b}
\end{equation}
where $c_i=c_i(r)$. The metric ansatz \eqref{g10b} is the most general one, given the $U(1)\times SU(2)\times SU(2)$
symmetries of the KW plasma \cite{kw}.
The frames $\{e_{\theta_a},e_{\phi_a}\}$ are defined so that the metric on a unit $T^{1,1}$ (the  angular 
part of the conifold) is\footnote{See \cite{aby} for explicit expressions.} 
\begin{equation}
\left(dT^{1,1}\right)^2=e_\psi^2+\sum_{a=1}^2
\left(e_{\theta_a}^2+e_{\phi_a}^2\right)\,.
\eqlabel{frames}
\end{equation}

For the dilaton we assume $\phi=\phi(r)$ and for the five-form 
\begin{equation}
F_5=\calf_5+\star\calf_5\,,\qquad \calf_5=-4\ dvol_{T^{1,1}}\,.
\eqlabel{5form}
\end{equation}
In \eqref{5form} the 5-form flux is chosen in such a way that $\gamma=0$ solution corresponds to $c_4=c_5=1$.
To leading order in $\gamma$ we further parameterize 
\begin{equation}
\begin{split}
c_1=&r \left(1-\frac{r_0^4}{r^4}\right)^{1/2} e^{-\ft 53 \nu_1}\left(1+a+4b\right)\,,\\
c_2=&r e^{-\ft 53 \nu_1}\,,\\
c_3=&\frac{1}{r\left(1-\frac{r_0^4}{r^4}\right)^{1/2}}\  e^{-\ft 53 \nu_1}\left(1+b\right)\,,\\
c_4=&e^{\nu_1-\nu_2}\,,\\
c_5=&e^{\nu_1+\nu_2}\,,
\end{split}
\eqlabel{defm}
\end{equation} 
where the asymptotic normalizable components of gravitational modes $\nu_1$ and $\nu_2$ are dual to
KW plasma operators of dimension-eight and dimension-six correspondingly \cite{aby}.

Equation of motion for $\nu_2$ determined to order $\calo(\gamma^2)$ from \eqref{aaa} takes form 
\begin{equation}
0=\nu_2''+\frac{5r^4-r_0^4}{r(r^4-r_0^4)}\ \nu_2'-\frac {12r^2}{r^4-r_0^4}\ \nu_2\,.
\eqlabel{nu2eom}
\end{equation}
The absence of non-normalizable mode asymptotically as $r\to \infty$ and the regularity at the horizon 
$r\to {r_0}_+$ determines unique solution to \eqref{nu2eom}:
\begin{equation}
\nu_2(r)=0\,.
\eqlabel{nu2zero}
\end{equation}
Vanishing of $\nu_2$ implies
that the remaining modes $\{a,b,\nu_1\}$ are  exactly the same\footnote{We explicitly verified this.}  as for the 
near-extremal D3 branes \cite{gkt1}
\begin{equation}
\begin{split}
a=&-\g\ \frac{15r_0^4}{2r^4}\left(25 \frac{r_0^4}{r^4}-79 \frac{r_0^8}{r^8}+25\right)\,,\\
b=&\g\ \frac{15r_0^4}{2r^4}\left(5 \frac{r_0^4}{r^4}-19 \frac{r_0^8}{r^8}+5\right)\,,\\
\nu_1=&\g\ \frac{15r_0^8}{32r^8}\left(1+\frac{r_0^4}{r^4}\right)\,.
\end{split}
\end{equation}
The dilaton $\phi$ also receives $\g$ corrections, $\phi\propto \gamma$ \cite{gkt1}. 
It is easy to see that to order $\calo(\g)$ gravitational perturbations do not mix with the dilaton 
perturbation; moreover to study gravitational perturbations we can consistency set $\phi=0$.
The Hawking temperature corresponding to the metric \eqref{g10b}
is \cite{gkt1}
\begin{equation}
T=T_0 \left(1+15\g\right)\equiv \frac{r_0}{\pi}\left(1+15\g\right)\,.
\eqlabel{temp}
\end{equation}

\section{Shear viscosity of KW plasma at finite 't Hooft coupling}
A relation between hydrodynamics of gauge theory plasmas and the spectrum of quasinormal modes in 
dual near-extremal string theory backgrounds is explained in \cite{ks}. Computation we need to perform 
here literally repeats the analysis done in \cite{f2} --- the only difference being that here we have a 
slightly more complicated background geometry \eqref{g10b}. Thus, we move directly to the results of the analysis.

A shear quasinormal mode
\begin{equation} 
Z_{shear}=Z_{shear,0}+\g\ Z_{shear,1}+\calo(\g^2) 
\eqlabel{skw}
\end{equation}
of frequency and momentum $(\omega,q=|\ \vec{q}\ |)$ in the background geometry \eqref{g10b}
satisfies the following equations
\begin{equation}
\begin{split}
0=&Z_{shear,0}''+\frac{x^2\qq^2+\ww^2}{x(\ww^2-x^2 \qq^2)}\ Z_{shear,0}'+\frac{\ww^2-x^2 \qq^2}{x^2(1-x^2)^{3/2}}\ 
Z_{shear,0}\,,\\
0=&Z_{shear,1}''+\frac{x^2\qq^2+\ww^2}{x(\ww^2-x^2 \qq^2)}\ Z_{shear,1}'+\frac{\ww^2-x^2 \qq^2}{x^2(1-x^2)^{3/2}}\ 
Z_{shear,1}+J_{shear,0}^{KW}\,,
\end{split}
\eqlabel{zsea1}
\end{equation}
where the source  $J_{shear,0}^{KW}$ is a functional of the zero's order shear mode $Z_{shear,0}$
\begin{equation}
\begin{split}
J_{shear,0}^{KW}=&\calc_{shear,KW}^{(4)}  \frac{d^4 Z_{shear,0}}{d x^4}+\calc_{shear,KW}^{(3)}
\ \frac{d^3 Z_{shear,0}}{d x^3}+\calc_{shear,KW}^{(2)}
\  \frac{d^2 Z_{shear,0}}{d x^2}\\
&+\calc_{shear,KW}^{(1)}\ \frac{d Z_{shear,0}}{d x}
+\calc_{shear,KW}^{(0)}\ Z_{shear,0}\\
\equiv&{\hat{\calc}}_{shear,KW}^{(1)}\ \frac{d Z_{shear,0}}{d x}
+{\hat{\calc}}_{shear,KW}^{(0)}\ Z_{shear,0}\,,
\end{split}
\eqlabel{sourceshear}
\end{equation}
where in the second equality we used the first equation in \eqref{zsea1}.
As in \cite{f2}, 
\begin{equation}
\ww=\frac{\w}{2\pi T_0}\,,\qquad \qq=\frac{q}{2\pi T_0}\,,
\eqlabel{wwqqdef}
\end{equation}
and instead of the radial coordinate $r$ we used 
\begin{equation}
x\equiv \left(1-\frac{r_0^4}{r^4}\right)^{1/2}\,.
\eqlabel{xdef}
\end{equation}
The connection coefficients $\calc_{shear,KW}^{(i)}$ are given explicitly in  Appendix A.

Introducing\footnote{The incoming wave boundary conditions must be imposed as explained in \cite{f4}.} 
\begin{equation}
Z_{shear}(x)=x^{-i\ww(1-15\gamma)} z_{shear}(x)\,,
\eqlabel{diffbh}
\end{equation}
a spectrum of the quasinormal modes is determined solving \eqref{zsea1} subject to the boundary 
conditions \cite{ks}
\begin{equation}
z_{shear}\bigg|_{x\to 0_+}=1\,,\qquad z_{shear}\bigg|_{x\to 1_-}=0\,.
\eqlabel{diffbh1}
\end{equation}
 
Superficially, the quasinormal modes of KW plasma are different from those of the $\caln=4$ SYM plasma.
Indeed, given that \cite{f2}
\begin{equation}
\begin{split}
\calc_{shear,\caln=4}^{(4)}=&45(1-x^2)^4
\end{split}
\eqlabel{csh4n4}
\end{equation}
is different from $\calc_{shear,KW}^{(4)}$ (see \eqref{csh4}), as well as all the other coefficients,
one is tempted to conclude that 
\[
J_{shear,0}^{KW}\ne J_{shear,0}^{\caln=4}\,.
\]
This is incorrect however. The source term in \eqref{zsea1} must be evaluated on the zero's order shear mode
$Z_{shear,0}$. Using the first equation in \eqref{zsea1} (which is a supergravity approximation and thus is the 
same for both KW and $\caln=4$ SYM plasmas) it is straightforward to show that even though 
$\calc_{shear,KW}^{(i)}\ne  \calc_{shear,\caln=4}^{(i)}$, 
 \begin{equation}
J_{shear,0}^{KW}\equiv J_{shear,0}^{\caln=4}
\eqlabel{main}
\end{equation}
for all $(\ww,\qq)$.

Eq.\eqref{main} is our main result. It establishes that the full spectrum of shear quasinormal modes is the same in KW and $\caln=4$ 
SYM plasmas\footnote{We believe the same statement is correct for the spectrum of sound quasinormal modes as well.}. To some extend 
the equivalence of the source terms for the KW and the $\caln=4$ CFTs could have been expected given that the scalar mode $\nu_2$ deforming the 
$U(1)$ fiber inside the $T^{1,1}$ of the dual KW geometry is not excited at $\calo(\gamma)$ order, see \eqref{nu2zero}. 
Thus, it is reasonable to expect 
that the Kaluza-Klein reduction of type IIB string theory on  $S^5$ would be the same as for the rigid $T^{1,1}$ (apart from the obvious rescaling 
of the five-dimensional gravitational coupling)\footnote{We expect a similar argument to apply to more general examples of $AdS_5\times SE^5/CFT_4$ 
correspondence.}. 

In order to relate to the analysis of the next section we reproduce from \cite{f2,f4} the lowest quasinormal mode 
of \eqref{zsea1} and its dispersion relation:
\begin{equation}
\begin{split}
z_{shear,0}=&z_{shear,0}^{(0)}+i\qq z_{shear,0}^{(1)}+\calo(\qq^2)\,,\\
z_{shear,1}=&z_{shear,1}^{(0)}+i\qq z_{shear,1}^{(1)}+\calo(\qq^2)\,,
\end{split}
\eqlabel{sss1}
\end{equation} 
\begin{equation}
\begin{split}
z_{shear,0}^{(0)}=&1\,,\qquad z_{shear,0}^{(1)}=\frac 12 \frac{\qq}{\ww}\ x^2\,,\\
z_{shear,1}^{(0)}=&\frac{25}{16}x^2\left(x^4-4x^2+5\right)\,,
\\
z_{shear,1}^{(1)}=&-\frac{1}{32\qq \ww}x^2\biggl(\qq^2\left(-240-1565 x^2-860 x^4+695x^6\right)\\
&+16\ww^2
\left(594-264x^2+43x^4\right)\biggr)-\frac{15\qq}{2\ww} x^2\,.
\end{split}
\eqlabel{sss4}
\end{equation}
Imposing the Dirichlet condition on $z_{shear,0}$ at the boundary determines the 
lowest shear quasinormal frequency
\begin{equation}
\ww=-i\ \Gamma_\eta\ \qq^2+\calo(\qq^3)\,,\qquad \Gamma_\eta=\frac 12 +\frac{105}{2}\g+\calo(\g^2)\,.
\eqlabel{shearres}
\end{equation}

\section{Shear viscosity in effective higher-derivative gauge/gravity models}
Computation of the quasinormal modes in near extremal backgrounds of 
ten dimensional type IIB supergravity including $\calo(\a'^3)$ 
higher-derivative corrections is technically rather involved. One difficulty is 
a high dimensionality of the background space-time. For the latter reason it is desirable to do the computations in
lower dimensional effective description. Of course, had a relevant Kaluza-Klein (KK) reduction of the higher-derivative 
type IIB supergravity been know, both ways to do the computations are equivalent. In the absence 
of a consistent KK reduction one usually resorts to guesses as to what the lower dimensional effective action 
would look like. In this section we present one example which illustrates pitfalls of such an approach.

Suppose we would like to guess an effective action corresponding to \eqref{aaa} KK reduced on $S^5$ (or $T^{1,1}$).
As a minimal requirement we would ask that the near extremal background of our effective action faithfully reproduces 
the thermodynamics of the full ten dimensional background. A natural guess  (satisfying  the minimal requirement)
would then be 
\begin{equation}
I=  \frac{1}{ 16\pi G_{5}} \int d^{5} x \sqrt {-g}
\ \bigg[ R  +12+ 
\  \gamma \  W \bigg]   \,,
\eqlabel{aaa5}
\end{equation}
where $W$ is given again by \eqref{rrrr}. 
Note that \eqref{aaa5} is precisely the effective action used in \cite{gkt1} to describe $\a'$  corrections 
of the near extremal D3 branes.

Repeating the analysis of the shear quasinormal modes in this case we find the same set of equations 
\eqref{zsea1}, albeit with the different source term:
\begin{equation}
\begin{split}
J_{shear,0}^{eff}=&\calc_{shear,eff}^{(4)}  \frac{d^4 Z_{shear,0}}{d x^4}+\calc_{shear,eff}^{(3)}
\ \frac{d^3 Z_{shear,0}}{d x^3}+\calc_{shear,eff}^{(2)}
\  \frac{d^2 Z_{shear,0}}{d x^2}\\
&+\calc_{shear,eff}^{(1)}\ \frac{d Z_{shear,0}}{d x}
+\calc_{shear,eff}^{(0)}\ Z_{shear,0}\\
\equiv&\hat{\calc}_{shear,eff}^{(1)}\ \frac{d Z_{shear,0}}{d x}
+\hat{\calc}_{shear,eff}^{(0)}\ Z_{shear,0}\,.
\end{split}
\eqlabel{sourceeff}
\end{equation}
The connection coefficients $\calc_{shear,eff}^{(i)}$ are given explicitly in  Appendix B.
Here, $J_{shear,0}^{eff}$ is actually different from $J_{shear,0}^{KW}$. Indeed, using the 
first equation in \eqref{zsea1} we find
\begin{equation}
\begin{split}
0\qquad \ne \qquad &J_{shear,0}^{KW}-J_{shear,0}^{eff}=-\frac{25 }{4}x(3 x^4-8 x^2+5)\  \frac{d Z_{shear,0}}{d x}\\
&-\frac{25}{4 (\qq^2 x^2-\ww^2)} 
\biggl(2 \qq^2 x^4 (3 x^2-4)-\ww^2 (9 x^4-16 x^2+5)\biggr)\ Z_{shear,0}\,.
\end{split}
\eqlabel{discr}
\end{equation}

We can explicitly compute the lowest quasinormal mode:
\begin{equation}
\begin{split}
z_{shear,0}^{(0)}=&1\,,\qquad z_{shear,0}^{(1)}=\frac 12 \frac{\qq}{\ww}\ x^2\,,\\
z_{shear,1}^{(0)}=&0\,,
\\
z_{shear,1}^{(1)}=&-\frac{x^2}{2 \qq \ww} \biggl(\qq^2 (45 x^6-60 x^4-90 x^2-15)+\ww^2 (-264 x^2+43 x^4+594)\biggr)\\
&-\frac{15\qq}{2\ww} x^2\,.
\end{split}
\eqlabel{sss4e}
\end{equation}
While \eqref{sss4e} is clearly different from \eqref{sss4}, nonetheless, it has the same dispersion relation 
\eqref{shearres}.

Given \eqref{discr}, the full shear quasinormal spectrum of $\caln=4$ SYM plasma is different  from the one corresponding to
the purported holographic dual to \eqref{aaa5}. The difference exists even for the lowest quasinormal mode,
although in higher orders in the hydrodynamic approximation.

\section{Conclusion}
In this letter we conjectured universality of the shear viscosity to entropy ratio of different 
CFT plasmas at finite 't Hooft coupling. While we presented only one example for a comparison with the $\caln=4$ SYM plasma,
namely that of  Klebanov-Witten plasma
and only to leading order in the inverse 't Hooft coupling, 
the fact that the full shear quasinormal spectra of both plasmas agree  suggests 
that the agreement extends to other CFT plasmas as well, and probably also beyond leading 
inverse 't Hooft coupling correction.  

Proving our conjecture to all orders in $\a'$ (or the inverse 't Hooft coupling) is unlikely to be possible.
We believe that at least to order $\calo\left(\lambda^{-3/2}\right)$ the universality can we established 
by carefully analyzing Kaluza-Klein reduction of  \eqref{aaa} on five dimensional Einstein-Sasaki manifolds. 
A progress on this issue can also be made by studying CFT plasmas in a boost-invariant setting, as
in \cite{f3}. 
We hope to report on this elsewhere. 

Within the supergravity approximation (equivalently at infinite 't Hooft coupling) the universality of the shear viscosity to the entropy
density ratio extends to gauge theory plasmas in various dimensions.  We conjectured here that the finite 't Hooft coupling corrections 
are universal for all four dimensional CFTs, which allow for a dual string theory description. It would be interesting to 
explore such corrections for conformal gauge theory plasmas in spacetime dimensions other than four.

\section*{Acknowledgments}
I would like to thank the organizers of ``String Theory - From Theory to Experiment'' workshop at Hebrew University of 
Jerusalem for hospitality where this work started. I would also like to thank the Galileo Galilei Institute 
for Theoretical Physics for the hospitality and the INFN for partial support during the completion of this work. 
My research at Perimeter Institute is supported in part by the Government
of Canada through NSERC and by the Province of Ontario through MRI.
I gratefully acknowledges further support by an NSERC Discovery
grant and support through the Early Researcher Award program by the
Province of Ontario.

\section*{Appendix}
\appendix

\section{Coefficients of $J_{shear,0}^{KW}$}
\begin{equation}
\begin{split}
\calc_{shear,KW}^{(4)}=&\frac{(1-x^2)^2}{\qq^2 x^2-\ww^2} \biggl(\qq^2 (45 x^6-90 x^4+57 x^2)-\ww^2 (37 x^4-74 x^2+49)\biggr)
\end{split}
\eqlabel{csh4}
\end{equation}
\begin{equation}
\begin{split}
\calc_{shear,KW}^{(3)}=&\frac{2 (x^2-1)}{(\qq^2 x^2-\ww^2)^2 x} \biggl(\qq^4 (315 x^{10}-585 x^8+261 x^6+57 x^4)
-8 x^2 \ww^2 \qq^2 (85 x^6\\
&-173 x^4+103 x^2-3)+\ww^4 (333 x^6-703 x^4+467 x^2-49)\biggr)
\end{split}
\eqlabel{csh3}
\end{equation}
\begin{equation}
\begin{split}
&\calc_{shear,KW}^{(2)}=-\frac{2}{x^2 (\qq^2 x^2-\ww^2) \sqrt{1-x^2}} \biggl(
\qq^4 (83 x^{10}-249 x^8+237 x^6-71 x^4)\\
&+2 \qq^2 \ww^2 x^2 (x^2-1) (25 x^4-50 x^2+37)+\ww^4 (1-x^2) (37 x^4-74 x^2+49)\biggr)\\
&+ \frac{1}{8 (\qq^2 x^2-\ww^2)^3 x^2} \biggl(\qq^6 x^6 (14015 x^8+3122 x^2-22940 x^6+5203 x^4+1368)\\
&-3 x^4 \ww^2 \qq^4 (16711 x^8-30396 x^6+10243 x^4+5594 x^2-1384)\\
&+ x^2 \qq^2 \ww^4 (-3864+60493 x^8-127764 x^6+75993 x^4-2554 x^2)-\ww^6 (22327 x^8\\
&-51324 x^6+38179 x^4-392-8022 x^2)\biggr)
\end{split}
\eqlabel{csh2}
\end{equation}
\begin{equation}
\begin{split}
&\calc_{shear,KW}^{(1)}=-\frac{2}{x^3 (\qq^2 x^2-\ww^2)^2 \sqrt{1-x^2}} \biggl(
\qq^6 x^6 (332 x^6-581 x^4+166 x^2+71)\\
&-3 x^4 \qq^4 \ww^2 (188 x^6-509 x^4+454 x^2-145)-3 x^2 \qq^2 \ww^4 (100 x^6-155 x^4+10 x^2+57)\\
&+\ww^6 (148 x^6-259 x^4+74 x^2+49)\biggr)\\
&-\frac{1}{8 (\qq^2 x^2-\ww^2)^4 x^3} \biggl(\qq^8 x^8 (8333 x^4+3122 x^2+1368-17580 x^6+5525 x^8)\\
&-2 x^6 \qq^6 \ww^2 (8582 x^2-5040-39148 x^6+25791 x^4+9983 x^8)+4 x^4 \qq^4 \ww^4 (-2148\\
&-30048 x^6+32523 x^4-6170 x^2+4943 x^8)+2 x^2 \qq^2 \ww^6 (816+17556 x^6-30025 x^4\\
&+11286 x^2+2935 x^8)-\ww^8 (-7980 x^6+2217 x^4+234 x^2+392+7105 x^8)\biggr)
\end{split}
\eqlabel{csh1}
\end{equation}
\begin{equation}
\begin{split}
&\calc_{shear,KW}^{(0)}=\frac{1}{8 x^4 (\qq^2 x^2-\ww^2) (1-x^2)^{3/2}} \biggl(\qq^4 x^6 (5621 x^2-1370-7540 x^4+3385 x^6)\\
&-2 x^2 \qq^2 \ww^2 (-7244 x^6+3917 x^4+1726 x^2-1568+3265 x^8)+\ww^4 (-1568-6692 x^6\\
&+4709 x^4+982 x^2+2665 x^8)\biggl)\\
&+ \frac{1}{4 x^4 (1-x^2) (\qq^2 x^2-\ww^2)} \biggl(
-4 \qq^6 x^6 (19 x^4-38 x^2+7)+4 x^4 \qq^4 \ww^2 (x^2+5) (x^2-7)\\
&+2 x^2 \qq^2 (25 x^6 (x^2-1) (3 x^2-4)+2 \ww^4 (91+55 x^4-110 x^2))-\ww^2 (4 \ww^4 (37 x^4-74 x^2\\
&+49)+25 x^4 (x^2-1) (9 x^4-16 x^2+5))\biggr)
\end{split}
\eqlabel{csh0}
\end{equation}

\section{Coefficients of $J_{shear,0}^{eff}$}
\begin{equation}
\begin{split}
\calc_{shear,eff}^{(4)}=&\frac{32(1-x^2)^4 (5 \qq^2 x^2-3 \ww^2)}{3 (\qq^2 x^2-\ww^2)} 
\end{split}
\eqlabel{cso4}
\end{equation}
\begin{equation}
\begin{split}
&\calc_{shear,eff}^{(3)}=\frac{64(x^2-1)^3}{3 x (\qq^2 x^2-\ww^2)^2}  \biggl(
5 \qq^4 x^4 (7 x^2+1)-2 x^2 \qq^2 \ww^2 (35 x^2-3)+3 \ww^4 (9 x^2-1)\biggr)
\end{split}
\eqlabel{cso3}
\end{equation}
\begin{equation}
\begin{split}
&\calc_{shear,eff}^{(2)}=-\frac{64 (x^2-1)^3}{3 (\qq^2 x^2-\ww^2) x^2 \sqrt{1-x^2}} \biggl(
7 \qq^4 x^4+5 x^2 \qq^2 \ww^2-3 \ww^4\biggr)\\
&+ \frac{x^2-1}{3 x^2 (\qq^2 x^2-\ww^2)^3} \biggl(20 \qq^6 x^6 (x^2-1) (307 x^4+107 x^2+24)-x^4 \qq^4 \ww^2 (21653 x^6\\
&-18675 x^4-4833 x^2+2080)+2 x^2 \qq^2 \ww^4 (12379 x^6-15043 x^4+2585 x^2+304)\\
&-3 \ww^6 (2399 x^6-3089 x^4+733 x^2+32)\biggr)
\end{split}
\eqlabel{cso2}
\end{equation}
\begin{equation}
\begin{split}
&\calc_{shear,eff}^{(1)}=-\frac{64 (1-x^2)^{3/2}}{3 (\qq^2 x^2-\ww^2)^2 x^3} 
\biggl(7 \qq^6 x^6 (4 x^2+1)-2 x^4 \qq^4 \ww^2 (23 x^2-18)\\
&-10 x^2 \qq^2 \ww^4 (3 x^2+1)+3 \ww^6 (4 x^2+1)\biggr)\\
&-\frac{1}{3 x^3 (\qq^2 x^2-\ww^2)^4} \biggl(20 \qq^8 x^8 (x^2-1) (145 x^6-254 x^4-83 x^2-24)-x^6 \qq^6 \ww^2 (7103 x^2\\
&-4160-34756 x^6+21582 x^4+9781 x^8)+x^4 \qq^4 \ww^4 (-896-49012 x^6+56106 x^4\\
&-15135 x^2+7587 x^8)+x^2 \qq^2 \ww^6 (448+4172 x^6-13530 x^4+4875 x^2+5385 x^8)\\
&-3 \ww^8 (-560 x^6-54 x^4+32+67 x^2+665 x^8)\biggr)
\end{split}
\eqlabel{cso1}
\end{equation}
\begin{equation}
\begin{split}
&\calc_{shear,eff}^{(0)}=-\frac{1}{3 (\qq^2 x^2-\ww^2) x^4 \sqrt{1-x^2}} \biggl(
15 \qq^4 x^6 (59 x^4+33-77 x^2)-2 x^2 \qq^2 \ww^2 (925 x^6\\
&+384-1019 x^4-65 x^2)+3 \ww^4 (101 x^2+375 x^6+128-529 x^4)\biggr)\\
&+\frac{32 (x^2-1)(\qq^2 x^2-\ww^2) (\qq^2 x^2+3 \ww^2)}{3 x^4} 
\end{split}
\eqlabel{cso0}
\end{equation}


\begin{thebibliography}{99}

\bibitem{m9711}
J.~M.~Maldacena,
Adv.\ Theor.\ Math.\ Phys.\  {\bf 2}, 231 (1998)
[Int.\ J.\ Theor.\ Phys.\  {\bf 38}, 1113 (1999)]
[arXiv:hep-th/9711200].

\bibitem{m2}
  O.~Aharony, S.~S.~Gubser, J.~M.~Maldacena, H.~Ooguri and Y.~Oz,
  Phys.\ Rept.\  {\bf 323}, 183 (2000)
  [arXiv:hep-th/9905111].

\bibitem{u1}
  A.~Buchel and J.~T.~Liu,
  Phys.\ Rev.\ Lett.\  {\bf 93}, 090602 (2004)
  [arXiv:hep-th/0311175].


\bibitem{u2}
  P.~Kovtun, D.~T.~Son and A.~O.~Starinets,
  Phys.\ Rev.\ Lett.\  {\bf 94}, 111601 (2005)
  [arXiv:hep-th/0405231].


\bibitem{u3}
  A.~Buchel,
  Phys.\ Lett.\  B {\bf 609}, 392 (2005)
  [arXiv:hep-th/0408095].

\bibitem{u4}
  P.~Benincasa, A.~Buchel and R.~Naryshkin,
  Phys.\ Lett.\  B {\bf 645}, 309 (2007)
  [arXiv:hep-th/0610145].

\bibitem{u5}
  D.~Mateos, R.~C.~Myers and R.~M.~Thomson,
  Phys.\ Rev.\ Lett.\  {\bf 98}, 101601 (2007)
  [arXiv:hep-th/0610184].


\bibitem{u55}
  K.~Landsteiner and J.~Mas,
  JHEP {\bf 0707}, 088 (2007)
  [arXiv:0706.0411 [hep-th]].


\bibitem{u6}
  A.~Buchel,
  ``Bulk viscosity of gauge theory plasma at strong coupling,''
  arXiv:0708.3459 [hep-th].

\bibitem{f1}
  A.~Buchel, J.~T.~Liu and A.~O.~Starinets,
  Nucl.\ Phys.\  B {\bf 707}, 56 (2005)
  [arXiv:hep-th/0406264].

\bibitem{f2}
  P.~Benincasa and A.~Buchel,
  JHEP {\bf 0601}, 103 (2006)
  [arXiv:hep-th/0510041].


\bibitem{f3}
  A.~Buchel,
  ``Shear viscosity of boost invariant plasma at finite coupling,''
  arXiv:0801.4421 [hep-th].


\bibitem{f4}
  A.~Buchel,
  ``Resolving disagreement for eta/s in a CFT plasma at finite coupling,''
  arXiv:0805.2683 [hep-th].


\bibitem{p1}
  Y.~Kats and P.~Petrov,
  ``Effect of curvature squared corrections in AdS on the viscosity of the dual
  gauge theory,''
  arXiv:0712.0743 [hep-th].


\bibitem{p2}
  M.~Brigante, H.~Liu, R.~C.~Myers, S.~Shenker and S.~Yaida,
  ``Viscosity Bound Violation in Higher Derivative Gravity,''
  arXiv:0712.0805 [hep-th].

\bibitem{gkt1}
  S.~S.~Gubser, I.~R.~Klebanov and A.~A.~Tseytlin,
  Nucl.\ Phys.\ B {\bf 534}, 202 (1998)
  [arXiv:hep-th/9805156].

\bibitem{pt1}
  J.~Pawelczyk and S.~Theisen,
  JHEP {\bf 9809}, 010 (1998)
  [arXiv:hep-th/9808126].


\bibitem{th1}
  A.~Buchel,
  Nucl.\ Phys.\  B {\bf 750}, 45 (2006)
  [arXiv:hep-th/0604167].


\bibitem{cor1}
  M.~T.~Grisaru and D.~Zanon,
  Phys.\ Lett.\ B {\bf 177}, 347 (1986).

\bibitem{cor2}
  M.~D.~Freeman, C.~N.~Pope, M.~F.~Sohnius and K.~S.~Stelle,
  Phys.\ Lett.\ B {\bf 178}, 199 (1986).


\bibitem{cor3}
  Q.~H.~Park and D.~Zanon,
  Phys.\ Rev.\ D {\bf 35}, 4038 (1987).

\bibitem{cor4}
  D.~J.~Gross and E.~Witten,
  Nucl.\ Phys.\ B {\bf 277}, 1 (1986).



\bibitem{hss}
  S.~de Haro, A.~Sinkovics and K.~Skenderis,
  Phys.\ Rev.\  D {\bf 68}, 066001 (2003)
  [arXiv:hep-th/0302136].

\bibitem{pt}
  G.~Policastro and D.~Tsimpis,
  Class.\ Quant.\ Grav.\  {\bf 23}, 4753 (2006)
  [arXiv:hep-th/0603165].


\bibitem{p}
  M.~F.~Paulos,
  ``Higher derivative terms including the Ramond-Ramond five-form,''
  arXiv:0804.0763 [hep-th].


\bibitem{lrw}
  H.~Liu, K.~Rajagopal and U.~A.~Wiedemann,
  JHEP {\bf 0703}, 066 (2007)
  [arXiv:hep-ph/0612168].

\bibitem{kw}
  I.~R.~Klebanov and E.~Witten,
  Nucl.\ Phys.\  B {\bf 536}, 199 (1998)
  [arXiv:hep-th/9807080].

\bibitem{aby}
  O.~Aharony, A.~Buchel and A.~Yarom,
  Phys.\ Rev.\  D {\bf 72}, 066003 (2005)
  [arXiv:hep-th/0506002].


\bibitem{ks}
  P.~K.~Kovtun and A.~O.~Starinets,
  Phys.\ Rev.\  D {\bf 72}, 086009 (2005)
  [arXiv:hep-th/0506184].

\end{thebibliography}
\end{document}